\documentclass[reprint, superscriptaddress, amsmath, amssymb, aps, prb, english]{revtex4-2}

\usepackage{graphicx}
\usepackage{enumerate}
\usepackage{dcolumn}
\usepackage{bm}
\usepackage{hyperref}
\usepackage[mathlines]{lineno}
\usepackage{siunitx}
\usepackage[version=4]{mhchem}
\hypersetup{colorlinks=true, urlcolor=blue, linkcolor=blue, filecolor=blue, citecolor=blue}

\begin{document}

\preprint{APS/123-QED}

\title{Optical memory in a MoSe$_2$/Clinochlore device}

\author{Alessandra Ames}
\affiliation{Departamento de Física, Universidade Federal de São Carlos, 13565-905, São Carlos, São Paulo, Brazil}
\author{Frederico B. Sousa}
\affiliation{Departamento de Física, Universidade Federal de São Carlos, 13565-905, São Carlos, São Paulo, Brazil}
\author {Gabriel A. D. Souza}
\affiliation{Departamento de Física, Universidade Federal de São Carlos, 13565-905, São Carlos, São Paulo, Brazil}
\author{Raphaela de Oliveira}
\affiliation{Brazilian Synchrotron Light Laboratory (LNLS), Brazilian Center for Research in Energy and Materials (CNPEM), 13083-100, Campinas, São Paulo, Brazil}
\author{Igor R. F. Silva}
\affiliation{Departamento de Física, Universidade Federal de São Carlos, 13565-905, São Carlos, São Paulo, Brazil}
\author{Gabriel L. Rodrigues}
\affiliation{Brazilian Nanotechnology National Laboratory (LNNano), Brazilian Center for Research in Energy and Materials (CNPEM), 13083-200, Campinas, São Paulo, Brazil}
\affiliation{“Gleb Wataghin” Institute of Physics, State University of Campinas, 13083-970 Campinas, São Paulo, Brazil}
\author{Kenji Watanabe}
\affiliation{Research Center for Electronic and Optical Materials, National Institute for Materials Science, 1-1 Namiki, Tsukuba 305-0044, Japan}
\author{Takashi Taniguchi}
\affiliation{Research Center for Materials Nanoarchitectonics, National Institute for Materials Science, 1-1 Namiki, Tsukuba 305-0044, Japan}
\author{Gilmar E. Marques}
\affiliation{Departamento de Física, Universidade Federal de São Carlos, 13565-905, São Carlos, São Paulo, Brazil}
\author{Ingrid D. Barcelos}
\affiliation{Brazilian Synchrotron Light Laboratory (LNLS), Brazilian Center for Research in Energy and Materials (CNPEM), 13083-100, Campinas, São Paulo, Brazil}
\author{Alisson R. Cadore}
\affiliation{Brazilian Nanotechnology National Laboratory (LNNano), Brazilian Center for Research in Energy and Materials (CNPEM), 13083-200, Campinas, São Paulo, Brazil}
\author{Victor López-Richard}
\affiliation{Departamento de Física, Universidade Federal de São Carlos, 13565-905, São Carlos, São Paulo, Brazil}
\author{Marcio D. Teodoro}
\email{\textcolor{black}{Corresponding author: mdaldin@ufscar.br}}
\affiliation{Departamento de Física, Universidade Federal de São Carlos, 13565-905, São Carlos, São Paulo, Brazil}

\date{\today}


\begin{abstract}

Two-dimensional heterostructures have been crucial in advancing optoelectronic devices utilizing van der Waals materials. Semiconducting transition metal dichalcogenide monolayers, known for their unique optical properties, offer extensive possibilities for light-emitting devices. Recently, a memory-driven optical device, termed a Mem-emitter, was proposed using these monolayers atop dielectric substrates. The successful realization of such devices heavily depends on selecting the optimal substrate. Here, we report a pronounced memory effect in a MoSe$_2$/clinochlore device, evidenced by electric hysteresis in the intensity and energy of MoSe$_2$ monolayer emissions. This demonstrates both population-driven and transition-rate-driven Mem-emitter abilities. Our theoretical approach correlates these memory effects with internal state variables of the substrate, emphasizing that clinochlore layered structure is crucial for a robust and rich memory response. This work introduces a novel two-dimensional device with promising applications in memory functionalities, highlighting the importance of alternative insulators in fabricating van der Waals heterostructures.
\end{abstract}


\maketitle



\section{Introduction}

The confinement of electrons in two dimensions has led to novel physical phenomena extensively studied over the past years in isolated atomically thin layers of van der Waals (vdW) materials~\cite{geim2009graphene,manzeli20172d,Mueller2018,Shree2020,Wang2018}. 
Among these phenomena, memory effects, such as resistive memory responses, in two-dimensional (2D) materials have garnered significant interest due to their potential computing functionalities~\cite{xue2022integrated,kumar2023optoelectronic,Zhang2022,Pal2023,CadoreAPL2016}.
Recently, a theoretical proposal introduced an emitting device with memory capabilities based on its exposure history to light or other stimuli, introducing a new class of devices termed Mem-emitters~\cite{lopezrichard2024}.

Semiconducting 2D transition metal dichalcogenides (TMDs) are promising candidates for the active medium of Mem-emitter devices due to their unique optical and electronic properties~\cite{manzeli20172d,Wang2018,CHOI2017,Mueller2018,Shree2020}.
TMD monolayers exhibit high carrier mobility~\cite{radisavljevic2011single}, a direct bandgap in the visible range~\cite{mak2010atomically,splendiani2010emerging,chaves2020bandgap}, strong many-body effects~\cite{mak2013tightly,chernikov2014exciton,ugeda2014giant,he2014tightly,sousa2023ultrafast,Timmer2024}, and tunable properties through doping~\cite{onofrio2017novel,sousa2024effects,Cadore2024LED}, strain~\cite{roldan2015strain,sousa2024disentangling,Montblanch2021,Cavalini2024,Costa2023,Serati2022}, and electric fields~\cite{chaves2020bandgap,Barbone2018,Nutting2021,Rosa2024,Feuer2023}, making them ideal for Mem-emitter applications~\cite{lopezrichard2024}.

Additionally, the performance of the TMD-based devices is strongly dependent on the interface between the active material and the insulator that separates it from the gate electrode~\cite{illarionov2020insulators,Geim2013,ingrid2023phyllosilicates,Costa2023,chaves2020bandgap}. Thus, significant efforts are being made to identify suitable insulating 2D materials that preserve the qualities of the 2D semiconductor in vdW heterostructures (vdWHS)~\cite{illarionov2020insulators,Geim2013,ingrid2023phyllosilicates}. Hexagonal boron nitride (hBN), the most extensively studied vdW insulator, is chemically inert and has a flat surface free of dangling bonds~\cite{zhang2017two,Geim2013,illarionov2020insulators}, making it widely used as a gate insulator. Despite its advantages, the high cost of 2D hBN poses an obstacle to its use in large-scale nanodevice fabrication, highlighting the need for alternative vdW insulators~\cite{ingrid2023phyllosilicates,Mahapatra2024}.

Phyllosilicates are naturally abundant vdW insulators that have recently emerged as promising materials for use in vdWHS~\cite{frisenda2020naturally,cadore2022exploring,ingrid2023phyllosilicates,Mahapatra2024,mania2017,Nutting2021,Prando2021,Gadelha2021,Barcelos2018,Castellanos-GomezSmall2011,RaphaNano2024}. Among these phyllosilicates, clinochlore is a layered crystal with the general chemical formula $\mathrm{Mg_5Al(Si_3Al)O_{10}(OH)_8}$, which, similarly to hBN, exhibits a wide bandgap~\cite{de2022high}, a flat surface over large areas~\cite{de2022high}, and a dielectric constant of $\sim$4.3~\cite{kawahala2024shaping}. However, clinochlore is a hydrated mineral~\cite{de2024water} that naturally contains impurities, point defects, and water nanoconfined in the vdW gap that might affect its insulating properties and interactions in a vdWHS~\cite{de2022high,de2024water}. In contrast to hBN, the properties of clinochlore when integrated with other layered materials still require thorough investigation.

Here, we report a robust optical memory effect in a vdWHS device based on a monolayer (1L) MoSe$_2$ atop ultrathin clinochlore flakes. We conducted photoluminescence (PL) spectroscopy experiments to examine the 1L-MoSe$_2$ light emission under an external voltage bias. A hysteresis in the intensity and energy of both exciton and trion emissions emerges when sweeping the gate voltage at an appropriate sweep rate. This effect can also be tuned by varying the sweep amplitude and excitation power. Notably, the negligible hysteresis loop in reference 1L-MoSe$_2$/hBN devices suggests that clinochlore layers play a crucial role in this optical memory effect. Calculations based on voltage-dependent polarization fluctuations and charge carrier population processes allowed us to correlate our experimental hysteresis with dynamic internal state variables. We experimentally confirm that excitons in 1L-TMDs act as naturally transition-rate driven mem-emitters, while trions exhibit a combination of both transition-rate and population-driven abilities, as reported in Ref.~\citenum{lopezrichard2024}. The theoretical outcomes indicate that disordered insulators are more suitable for introducing optical memory effects, corroborating the rich memory response observed in our clinochlore-based device. Thus, our results demonstrate the experimental realization of a Mem-emitter device based on a vdWHs of a 1L-TMD on clinochlore layers, highlighting the potential of phyllosilicates as gate insulators in future nanotechnological devices.


\section{Results and Discussion}

\begin{figure}
    \centering
    \includegraphics{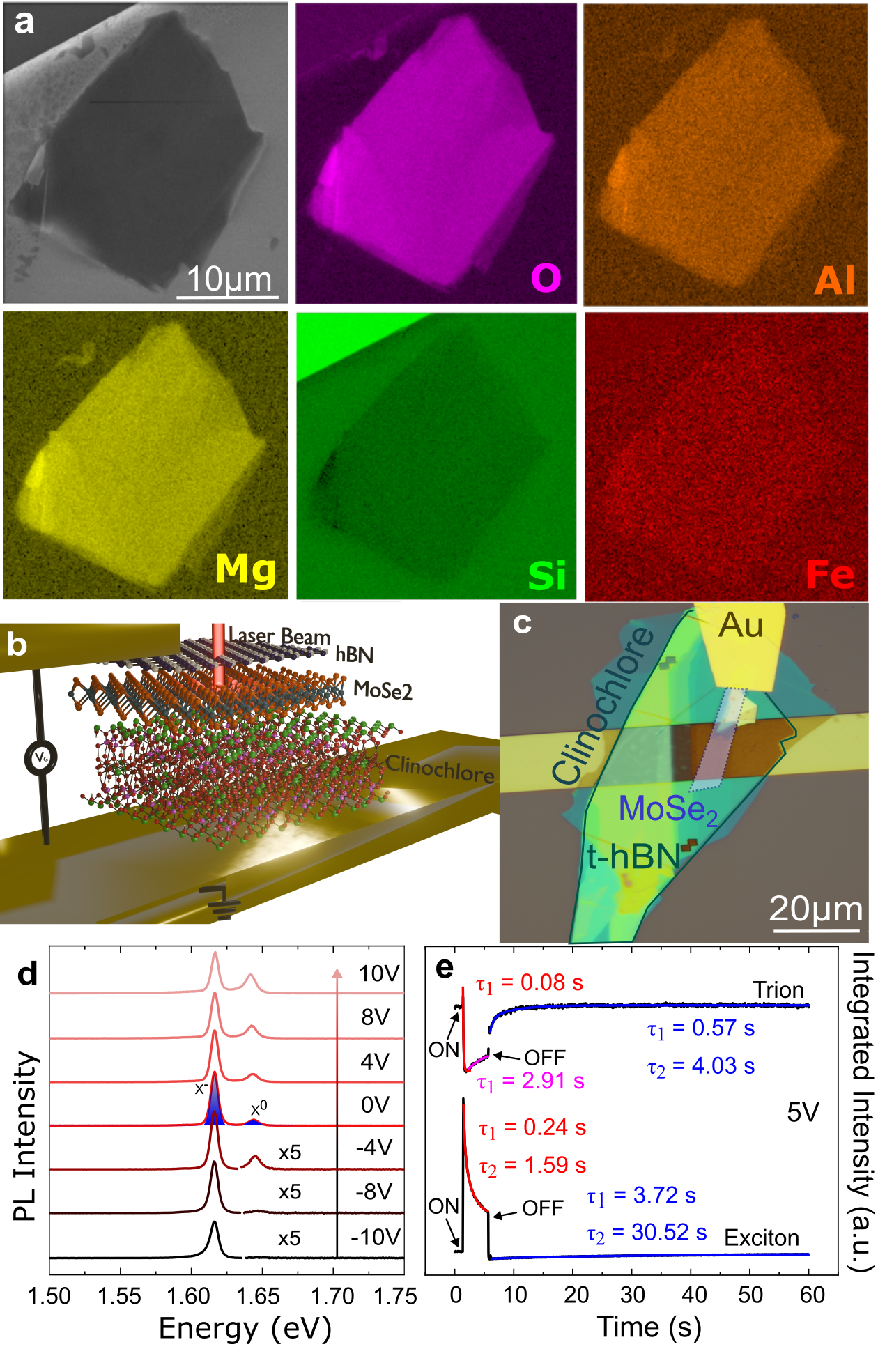}
    \caption{\textbf{Clinochlore-based MoSe$_2$ device.}
    \textbf{a} Clinochlore SEM image and EDS chemical maps of its constituent elements in addition to Fe impurity. 
    \textbf{b,c} Schematic view (b) and optical microscopy image (c) of our t-hBN/1L-MoSe$_2$/clinochlore/Au-electrode device. 
    \textbf{d} PL spectra of 1L-MoSe$_2$ acquired in the clinochlore sample at different gate voltages ranging from $-$10~V to 10~V. X$^0$ and X$^-$ emissions are denoted in the PL spectrum at 0~V.
    \textbf{e} Time-resolved PL emission transients recorded for the integrated intensity of excitons and trions in the 1L-MoSe$_2$ under the application of 5-second rectangular voltage pulses with a 5V amplitude. Each spectrum was taken over 0.03 seconds.}
    \label{fig1}
\end{figure}

The natural clinochlore crystal used in this study was pre-characterized using several experimental techniques detailed elsewhere~\cite{de2022high,de2024water}. Here, we exfoliated our natural sample onto a Si/Au (100 nm) substrate and performed hyperspectral mapping using Energy-Dispersive Spectroscopy (EDS) to confirm the homogeneous composition of the obtained flakes before the sample fabrication process. The scanning electron microscopy (SEM) image of clinochlore with its elemental EDS maps is shown in Figure~\ref{fig1}a. We observe a homogeneous distribution of the expected elements Mg, O, Si, and Al, along with Fe impurities typically present in natural clinochlore samples. The low chemical contrast for Fe impurities is related to its concentration (about 6$\%$wt~\cite{de2022high}), which is close to the detection limit of the EDS resolution (approximately 2-5$\%$). The low contrast for the Si map is due to the overlap with the Si content of the substrate.

After confirming the homogeneity of the exfoliated clinochlore sample composition, we proceeded with the sample fabrication. Figure~\ref{fig1}b presents a schematic view of our clinochlore-based device in a capacitor-like structure. The 1L-MoSe$_2$ is sandwiched between the top thin (t) hBN and bottom clinochlore flakes. Electrical contact from the Au electrode to the 1L-MoSe$_2$ was achieved through a few-layer graphite (FLG), with a voltage bias applied between the 1L-MoSe$_2$ and the bottom Au electrode (grounded contact). See Methods and Section~S1 for further details regarding sample fabrication and the insulating behavior of clinochlore, respectively. An optical microscope image of our fabricated hBN/FLG/1L-MoSe$_2$/clinochlore/Au-electrode device, with each part highlighted, is shown in Figure~\ref{fig1}c. Additionally, a similar sample with the bottom layer formed by hBN instead of clinochlore was fabricated as a reference device (Figure~S2). Hereafter, the samples will be referred to as the clinochlore and reference samples.

Figure~\ref{fig1}d shows the 1L-MoSe$_2$ PL spectra obtained from the clinochlore sample at 3.6~K, with gate voltages ranging from $-$10~V to 10~V. Emissions from neutral (X$^0$) and charged (X$^-$, trion) excitons are marked in the PL spectrum at 0V, centered at 1.645~eV and 1.616~eV, with full widths at half maximum (FWHM) of 8.9~meV and 6.7~meV, respectively. The corresponding PL spectra of the reference sample, shown in the Supporting Information (Figure~S2), exhibit emission energies (FWHM) of 1.644~eV (4.3~meV) and 1.617~eV (4.0~meV) for X$^0$ and X$^-$ peaks at 0~V, respectively. These energy and FWHM values were extracted by fitting the spectra to two Lorentzian peaks, aligning with the values reported in the literature for hBN-encapsulated 1L-WSe$_2$~\cite{cadiz2017excitonic}.
These results suggest that clinochlore is a viable alternative 2D insulator for emitting devices, despite its heterogeneities. 

Regarding the gate-dependent PL of the clinochlore sample, Figure~\ref{fig1}d illustrates a blueshift and an increase in intensity for the X$^0$ emission with increasing gate voltage. In contrast, the X$^-$ emission shows a blueshift and a relative decrease in intensity with voltage. Since the ground electrode is located at the bottom of the device, a positive gate voltage removes electrons from the 1L-MoSe$_2$, while a negative gate voltage injects electrons. This evolution of the PL spectra highlights the negative character of the charged exciton, with the X$^-$ emission dominating the spectra at negative voltages~\cite{mak2013tightly,ross2013electrical,shang2015observation,li2018revealing}.

An external electric field can modify internal state variables like polarization or charge carrier populations. For an emitting device based on a 1L-TMD atop a dielectric substrate, these changes impact the optical properties of the TMD~\cite{lopezrichard2024}. The dynamics of these internal state variable modifications create the conditions necessary for observing memory effects~\cite{lopezrichard2024}. The most straightforward way to prove the existence of non-equilibrium processes along with their proper time scales is through the analysis of time-resolved transients of the observables under constant bias. 

To examine transient processes, we investigated the temporal dependence of the 1L-MoSe$_2$ PL spectrum obtained from the clinochlore sample by continuously acquiring several PL spectra over 60 seconds, as shown in Figure~\ref{fig1}e. This measurement began at 0~V, and after 1 second, a gate voltage of 5~V was applied, resulting in an abrupt decrease (increase) in the X$^-$ (X$^0$) emission intensity. At 5.5 seconds, the voltage was turned off again.

For our 1L-MoSe$_2$ device, the optical properties are influenced by fluctuations in the substrate's polarization and carrier population under an applied voltage. We have then correlated a recent theoretical model that outlines the electric hysteresis of optical emission intensity and energy for 2D devices~\cite{lopezrichard2024} with our experimental data.

The dynamics of the nonequilibrium carrier population are governed by independent relaxation mechanisms, each characterized by relaxation times $\tau_i^{(n)}$\cite{lopezrichard2024}. The evolution of the fluctuation $\delta n_i$ in the carrier population responds as~\cite{lopezrichard2024}
\begin{equation}
    \frac{d(\delta n_i)}{dt} = - \frac{\delta n_i}{\tau_{i}^{(n)}} + g_{i}^{(n)}(V),
    \label{dn}
\end{equation}
where $n = n_0 + \sum_i \delta n_i$ represents the carrier population fluctuating around certain equilibrium values. Similarly, polarization fluctuations $P = P_0 + \sum_j \delta P_j$ follow an analogous dynamic, described by
\begin{equation}
   \frac{d(\delta P_j)}{dt} = - \frac{\delta P_j}{\tau_{j}^{(p)}} + g_{j}^{(p)}(V),
   \label{dp}
\end{equation}
with $\tau_{j}^{(p)}$ denoting the relaxation times for polarization dynamics. Here, $g_{i}^{(n)}(V)$ and $g_{j}^{(p)}(V)$ are the non-equilibrium carrier and polarization transfer functions, respectively, both dependent on the external gate voltage, $V$~\cite{lopezrichard2022,lopezrichard2024a}.
According to Ref.~\citenum{lopezrichard2024}, the emission energies of both excitons and trions evolve proportionally to $\varepsilon_0 V/d_2 + \sum_j \delta P_j (t)$, $d_2$ being the width of the dielectric substrate. Similarly, the exciton intensity follows this relationship, while the trion intensity is proportional to the amount of the electrons, $n_0 + \sum_i \delta n_i(t)$~\cite{Lundt2018}.

The coexistence of dynamics with contrasting relaxation times was confirmed through exponential decay fits using the solutions of Eqs.~\ref{dn} and~\ref{dp}~\cite{lopezrichard2024}, as displayed in Figure~\ref{fig1}e. Hence, these memory phenomena are governed by processes spanning a wide temporal scale, from fractions of a second to minutes.

\begin{figure}
    \centering
    \includegraphics{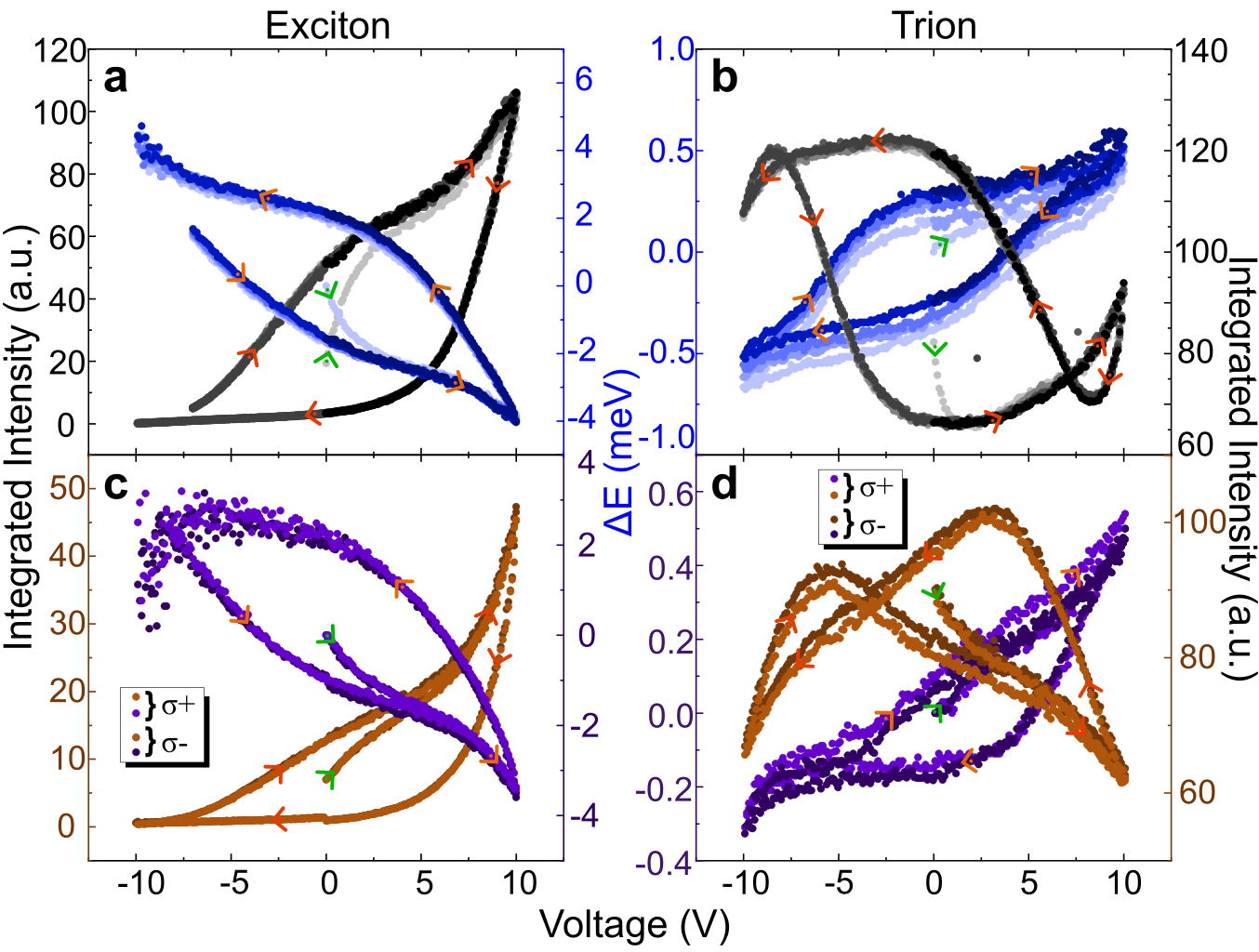}
    \caption{\textbf{Hysteresis effect in the PL emission of the clinochlore device under external voltage sweeps of 18 minutes period.}
    \textbf{a,b} Integrated intensity (in shades of black) and energy shift (in shades of blue) of X$^0$ (a) and X$^-$ (b) emissions as a function of the gate voltage, which was subsequently swept 5 times in cycles ranging from $-$10~V to 10~V. Each voltage cycle is presented in a different shade of black or blue for the intensity or energy data, respectively. 
    \textbf{c,d} The integrated intensity (in shades of brown) and energy shift (in shades of purple) of X$^0$ (c) and X$^-$ (d) emissions, after defect passivation, are shown as a function of the gate voltage. These measurements were detected at right ($\sigma^+$) and left ($\sigma^-$) circular polarizations. The voltage cycle for each detection is presented in a different shade of brown or purple for the intensity or energy data, respectively. For all graphs, green arrows indicate where the measurement initiates and orange arrows denote the direction of the voltage sweep. The energy shift $\Delta$E is relative to the first emission energy measured at 0~V.}
    \label{fig2}
\end{figure}

To reveal the memory effects of the device, using the slowest dynamics as a reference, we obtained 1L-MoSe$_2$ PL spectra from the clinochlore sample by sweeping the gate voltage (0~V → 10~V → $-$10~V → 0~V) with steps of 0.1~V and a cycling period $T$ until stable closed cycles were achieved. From these spectra, we determined the intensity (in black) and energy (in blue) of X$^0$ and X$^-$ emissions as a function of the gate voltage, as shown in Figures~\ref{fig2}a and \ref{fig2}b.

A Mem-emitter response emerges for both emissions in intensity and energy when sweeping the voltage. The hysteresis of the X$^0$ emission, displayed in Figure~\ref{fig2}a, presents a leaf shape with no crossings and a clockwise (counter-clockwise) loop direction for the intensity (energy). In contrast, the hysteresis of the X$^-$ emission, in Figure~\ref{fig2}b, exhibits a bow-like topology with two crossing points at opposite voltages for intensity concomitant with a leaf shape for energy, and a counter-clockwise (clockwise) loop direction for the intensity (energy) around V=0~V.

Figures~\ref{fig2}a and~\ref{fig2}b demonstrate that the hysteresis loops are consistent across five measured cycles. Their reproducibility is further confirmed in Figure~S3, which shows similar hysteresis loops for an experiment conducted under the same conditions but with the gate voltage cycle starting in the opposite direction. Additionally, we performed the hysteresis measurements detecting emissions at right ($\sigma^+$) and left ($\sigma^-$) circular polarizations to probe emissions from the K and K' valleys, respectively, as shown in Figures\ref{fig2}c and~\ref{fig2}d. The comparable results for both polarizations indicate that the hysteresis is not valley-dependent. Additionally, a contrast can be observed in the hysteresis loop shape for the trion intensity between Figures~\ref{fig2}b (bow-like) and \ref{fig2}d (butterfly-like). This evolution occurred after subjecting the sample to lengthy and continuous voltage scans of Figures~\ref{fig2}a and \ref{fig2}b and can be attributed to the passivation of specific activation channels for non-equilibrium electrons in the clinochlore layer. Furthermore, to examine the dependence of the memory effect on the clinochlore substrate, we conducted the same experiment on a reference sample. As shown in Figure~S2, the reference sample exhibited negligible hysteretic response. 
This observation aligns with previous reports of minor energy hysteresis for X$^0$ emission in hBN-encapsulated 1L-MoS$_2$ devices~\cite{roch2018quantum}. Recently, devices utilizing 1L-MoSe$_2$~\cite{choi2024tuning} and 1L-MoS$_2$~\cite{pucher2024strong} atop a perovskite substrate demonstrated significant hysteresis in the X$^-$/X$^0$ intensity ratio, which was attributed to a remanent polarization induced by the ferroelectric properties of the perovskite and underlines the fundamental role of the substrate in the hysteretic responses of such devices. We observe a similar hysteresis for the X$^-$/X$^0$ intensity ratio in the clinochlore sample (see Figure~S4). However, here we focus on the individual memory effects for X$^0$ or X$^-$ emissions that were not reported in Refs.~\citenum{choi2024tuning,pucher2024strong} and enable a more detailed comprehension of the distinct transient processes that govern these phenomena~\cite{lopezrichard2024}.

\begin{figure*}
    \centering
    \includegraphics{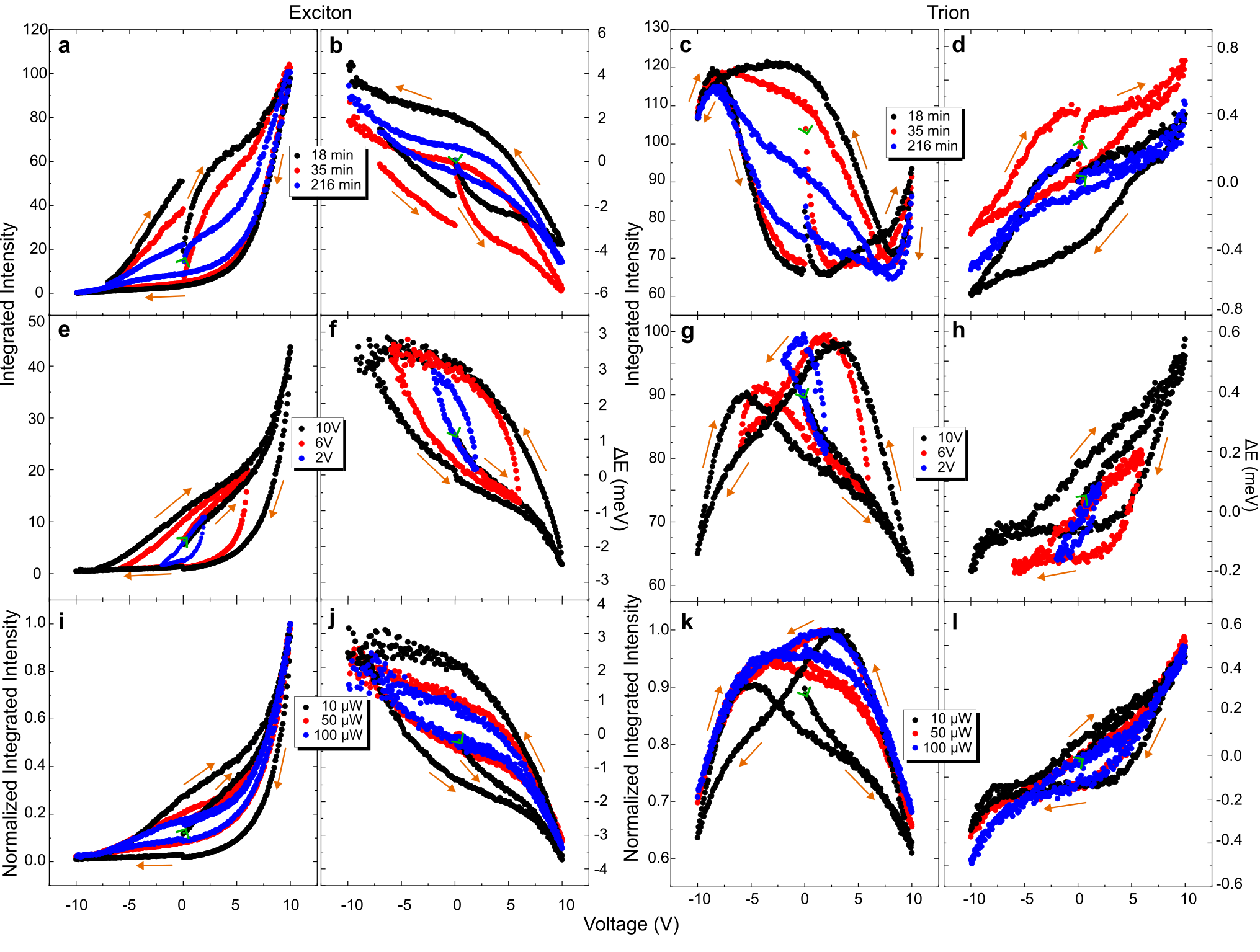}
    \caption{\textbf{Hysteresis effect in the PL emission of the clinochlore device at different conditions of voltage sweep rate, voltage sweep amplitude, and excitation power.}
    \textbf{a-d} Integrated intensity (a,c) and energy shift (b,d) of X$^0$ (a,b) and X$^-$ (c,d) emissions for different overall times of the gate voltage cycle before defect passivation.
    \textbf{e-h} Integrated intensity (e,g) and energy shift (f,h) of X$^0$ (e,f) and X$^-$ (g,h) emissions for different values of V$_{\mathrm{max}}$.
    \textbf{i-l} Normalized integrated intensity (i,k) and energy shift (j,l) of X$^0$ (i,j) and X$^-$ (k,l) emissions for different excitation powers.
    For all graphs, green arrows indicate where the measurement initiates and orange arrows denote the direction of the voltage sweep. The energy shift $\Delta$E is relative to the first emission energy measured at 0~V. The voltage sweeps consist of a single cycle as follows: 0~V $\rightarrow$ V$_{\mathrm{max}}$ $\rightarrow$ $-$V$_{\mathrm{max}}$ $\rightarrow$ 0~V. The acquisition time for each PL spectrum was maintained at 0.2 seconds, with the overall time varied by adjusting the delay time between each spectrum.}
    \label{fig3}
\end{figure*}

To gain deeper insights into the hysteresis effect in the clinochlore device, we acquired PL spectra by sweeping the gate voltage in a single cycle (0~V → V${\mathrm{max}}$ → $-$V${\mathrm{max}}$ → 0~V) under various conditions: different sweep rates, sweep amplitudes, and excitation powers, as displayed in Figure~\ref{fig3}. Figures~\ref{fig3}a-d present the intensity and energy of X$^0$ and X$^-$ emissions as a function of the gate voltage (V$_{\mathrm{max}}$ = 10~V), with measurements taken over periods of 18 minutes, 35 minutes, and 216 minutes. The results show that the hysteresis effect is more pronounced for shorter periods (black symbols) and that, according to Ref.~\citenum{lopezrichard2024}, will be closer to the optimal memory response.
Additionally, varying the sweep rate causes an energy shift in both X$^0$ and X$^-$ emissions across all gate voltages. We observe a redshift (blueshift) for the X$^0$ (X$^-$) emission when increasing the cycle time from 18 minutes to 35 minutes, with the energies for the 216-minute measurement falling between those of the faster cycles. Figures~\ref{fig3}e-h show the hysteresis in intensity and energy of X$^0$ and X$^-$ emissions for V${\mathrm{max}}$ = 10~V (black), 6~V (red), and 2~V (blue) after defect passivation.
Only the shape of the trion intensity fluctuation has been qualitatively modified.
The hysteresis amplitude in both intensity and energy increases with higher V${\mathrm{max}}$. The excitation power dependence of the hysteresis effect is displayed in Figures~\ref{fig3}i-l, showing normalized intensity and energy of X$^0$ and X$^-$ emissions. The hysteresis magnitude is largest at the lowest excitation power (10~$\mu$W) and exhibits similar reduced amplitudes at 50~$\mu$W and 100~$\mu$W.

\begin{figure*}
    \centering
    \includegraphics{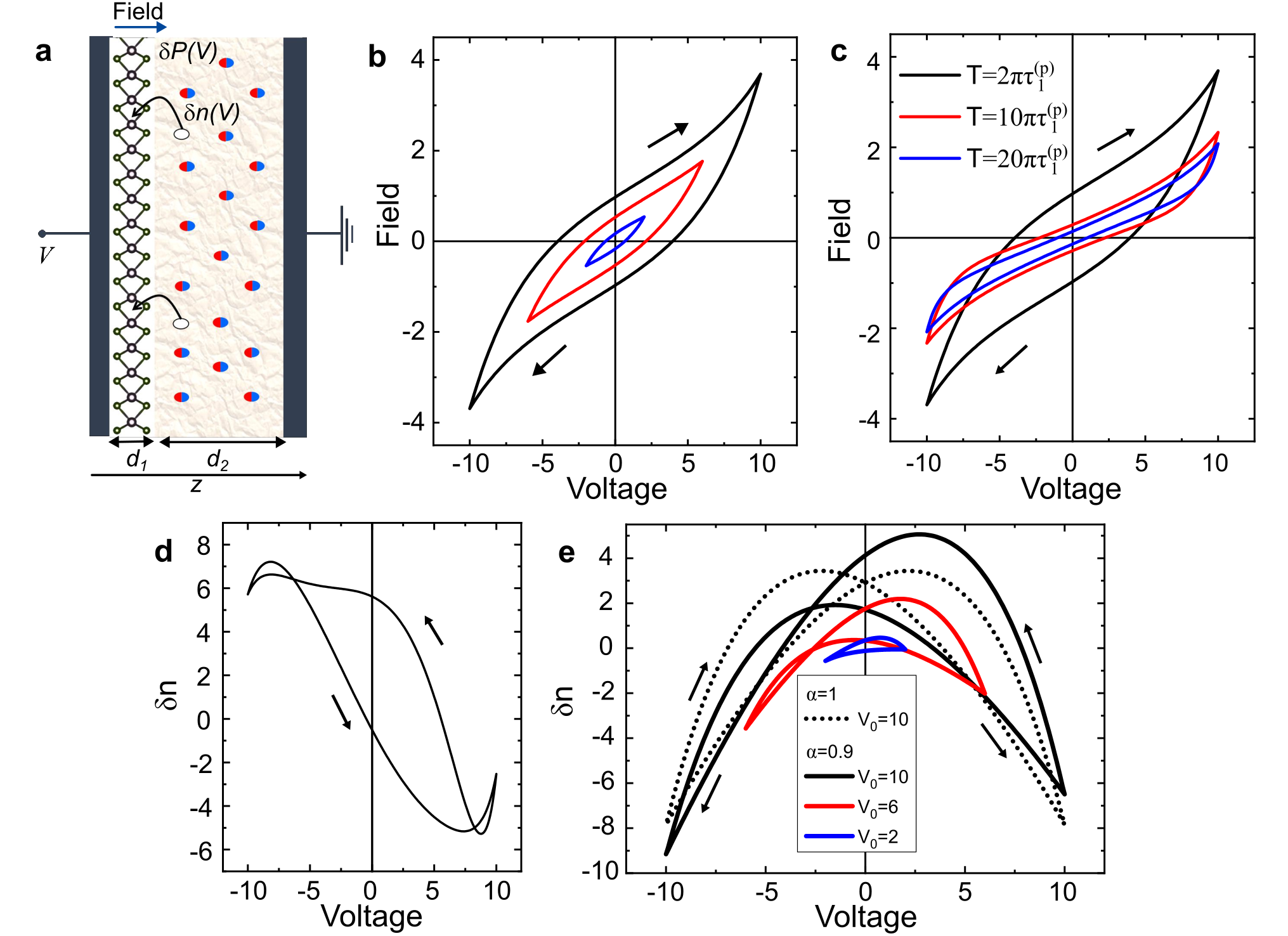}
    \caption{\textbf{Electric hysteresis modeling of polarization and carrier population fluctuations in a 1L-TMD-based device.} \textbf{a} Representation of a gated device composed by a 1L-TMD on a dielectric substrate. Polarization ($\delta P$) and charge carrier ($\delta N$) fluctuations occur in the substrate under an external electric field, impacting the optical properties of the TMD. Local field ($F_{\mathrm{TMD}}$) hysteresis considering polarization leakage channels and: \textbf{b}  increasing voltage amplitudes, \textbf{c} increasing voltage sweep periods. \textbf{d} Charge carrier fluctuation ($\delta N$) hysteresis considering five trapping and release channels. \textbf{e} Charge carrier fluctuation hysteresis for different voltage amplitudes. The arrows in (\textbf{b-e}) denote the direction of the voltage sweep.}
    \label{fig4}
\end{figure*}

The topology of a hysteresis loop is shaped by concurrent processes involving the device's internal state variables responding to external stimuli~\cite{lopezrichard2024}. These include fluctuations in the carrier population, $\sum_i \delta N_i$, or polarization, $\sum_j \delta P_j$, as described by Eqs.~\ref{dn} and ~\ref{dp}, respectively. Both processes are illustrated in panel~\ref{fig4}a. In the presence of carrier (dipoles) leakage~\cite{lopezrichard2024, WEI200564, JIN2023}, the electric field within the 1L-TMD, $F_{TMD}\propto \varepsilon_0 V/d_2 + \sum_j \delta P_j (t)$, can be calculated to follow the clockwise hysteresis pattern shown in panel~\ref{fig4}b, with the loop size decreasing as the voltage amplitude is reduced. The transfer functions used for this calculation are shown in Figure~S5a of the Supporting Information.

According to this model and Figures~\ref{fig2}a and \ref{fig2}b, it is evident that the transition energies for excitons and trions vary proportionally to $-F_{TMD}$ and $F_{TMD}$, respectively. The exciton transition rate should also change proportionally to $F_{TMD}$. Thus, the excitons contribute a transition rate-driven character to the mem-emitter capabilities of the device. Consistent with the observed collapse of the memory response with an increasing sweep period, as shown in Figures~\ref{fig3}a-d, the calculated hysteresis also shrinks in Figure~\ref{fig4}c, reinforcing the existence of optimal driving conditions.

The trion emission, however, shows a distinct separation between energy shifts and intensity fluctuations both before and after defect passivation, as depicted in Figures~\ref{fig2} and~\ref{fig3}. To reproduce the bow-like shape analogous to the observed experimental results in Figures~\ref{fig2}b and \ref{fig3}c, obtained before defect passivation, Figure~\ref{fig4}d illustrates the $\sum_i \delta n_i$ hysteresis calculated by combining five processes related to charge carrier population with contrasting relaxation times (assuming concomitant dynamics as detected in Figure~\ref{fig1}e).
Three of these processes are in the quasi-stationary regime ($\tau_i^{(n)} \ll T$), while the other two have relaxation times comparable to the sweep period ($\tau_i^{(n)} \sim T$). Their respective carrier transfer functions, $g_{i}^{(n)}(V)$, are displayed in Figure~S5b.

To reproduce the butterfly-like shape of the trion emission intensity after defect passivation, we consider two remaining dynamics with contrasting relaxation times and a slight inversion asymmetry, as shown in Figure~\ref{fig4}e (see the passivated transfer functions in Figure~S5c of the SI). The topology of this shape varies with the reduction of voltage amplitude, analogous to the experimental response observed in Figure~\ref{fig3}g. Thus, as anticipated in Ref.~\citenum{lopezrichard2024}, trions can exhibit a hybrid response, providing population-driven Mem-emitters abilities while also displaying transition rate-driven features, similar to excitons.
Hence, our modeled hysteresis loops suggest that for complex topologies multiple non-stationary fluctuations should be considered. Consequently, devices with a defect-rich dielectric substrate are more suited for a pronounced memory effect. This aligns with our experimental findings, where the clinochlore-based device, which contains significant Fe and H$_2$O impurities~\cite{de2022high,de2024water}, exhibits more complex and robust hysteresis compared to the hBN-based device.


\section{Conclusions}

In summary, we presented experimental evidence of a Mem-emitter device based on a 1L-MoSe$_2$ stacked on a few layers of clinochlore. We employed voltage-dependent PL experiments at low temperatures to investigate how this phyllosilicate affects the light emission of the TMD and to search for optical memory effects. Similar to the hBN-encapsulated 1L-MoSe$_2$, the 1L-MoSe$_2$/clinochlore device exhibited X$^0$ and X$^-$ emissions with narrow linewidths, demonstrating its potential as an alternative 2D insulator for use in emitting vdW heterostructures.
Additionally, the 1L-MoSe$_2$/clinochlore device showed robust hysteresis in the intensity and energy of X$^0$ and X$^-$ MoSe$_2$ emissions when an external electric field was applied, indicating a strong optical memory response. This contrasts with the weak hysteresis observed in the 1L-MoSe$_2$/hBN reference device. The temporal dependence of the emission of the 1L-MoSe$_2$/clinochlore device after a voltage pulse indicated distinct dynamical processes related to the internal state variables of the clinochlore substrate. These time-dependent processes govern the hysteresis of the optical observables. Therefore, the impurities in the clinochlore substrate introduce internal state variable processes, needed for the device's optical memory effect. Consequently, our results pave the way for diverse memory functionalities based on TMD-based 2D vdW heterostructures, highlighting the fundamental contribution of the clinochlore substrate to the reported Mem-emitter device.

\section{Methods}

\subsection{Sample Preparation}

Two sets of gated devices (clinochlore and reference samples) were fabricated following a similar fabrication process that will be described in the sequence. All ultrathin flakes were obtained from mechanical exfoliation of bulk crystals by applying the standard scotch tape technique. The natural clinochlore crystal was extracted from the geological environment of Minas Gerais - Brazil and characterized by energy dispersive spectroscopy (EDS) using a Helios 5 PFIB CXe DualBeam microscope. The MoSe$_2$ crystal was purchased from 2D Semiconductors, while the hBN crystal was grown by the temperature-gradient method~\cite{hBN}. The clinochlore (or hBN) crystal was mechanically exfoliated onto polydimethylsiloxane (PDMS) stamps and the selected flakes were dry transferred to pre-patterned Ti/Au (5/35 nm) electrodes previously fabricated by nanofabrication techniques~\cite{Feres2023}. The clinochlore (hBN) thickness selected for device fabrication was $\sim$30-50~nm to avoid leakage current during the voltage bias application in the capacitor-like structure~\cite{mania2017}. To create the gated hBN/1L-MoSe$_2$/clinochlore(hBN)/Au devices, we used the method described in Ref.~\cite{Cadore2024LED} and a commercial transfer system from HQ Graphene. The heterostructures composed by hBN/1L-MoSe$_2$/clinochlore(hBN)/Au were fabricated as follows: i) we first picked the top hBN ($\sim$10 nm) flake up from a SiO$_2$/Si substrate with a polycarbonate (PC) membrane onto a PDMS stamp at 90°C; ii) a few-layer graphite (FLG) flake was then picked up with the PC/hBN stamp at 60°C; iii) the PC/hBN/FLG stamp was aligned to the edge of the 1L-MoSe$_2$ flake, and the PC/hBN/FLG/1L-MoSe$_2$ was again picked up from the SiO$_2$/Si substrate at 60°C; iv) finally, PC/hBN/FLG/1L-MoSe$_2$ was aligned to the clinochlore (or hBN) flake already transferred to the Ti/Au electrode and brought into contact at 180°C, whereby PC adheres to the substrate, allowing PDMS to be peeled away, leaving PC/hBN/FLG/1L-MoSe$_2$/clinochlore(hBN) on Au electrode. PC is then dissolved in chloroform for $\sim$30 min at room temperature, leaving hBN/FLG/1L-MoSe$_2$/clinochlore(hBN) on the Au electrode. After heterostructure assembly, an additional Ti/Au (5/50nm) electrode was fabricated to contact the FLG flake and induce charge modulation to the MoSe$_2$ layer by an external voltage bias in a capacitor-like design.

\subsection{Optical Characterization}

For the optical characterization, clinochlore and reference samples were placed in a helium closed cycle cryostat (Attocube / Attodry 1000) at 3.6~K and were electrically connected to a chip carrier for the application of the external voltage bias that was controlled by a sourcemeter (Keithley 2400). The samples were excited by a linearly polarized laser beam with an excitation wavelength of 660 nm (Toptica - Ibeam). An aspheric lens (NA = 0.68) was used to focus the laser on the sample, and the backscattered PL signal was collected by the same lens and detected in a spectrometer equipped with a sensitive CCD camera (Andor / Shamrock - Idus). For the circular polarization measurements, the emitted signal was filtered by a linear polarizer and a quarter-wave plate.

\begin{acknowledgments}

This study was financed in part by the Fundação de Amparo à Pesquisa do Estado de São Paulo (FAPESP, Grant 2022/10340-2 and 2023/10905-2). 
The authors acknowledge the Brazilian Nanotechnology National Laboratory (LNNano) and Brazilian Synchrotron Light Laboratory (LNLS), both part of the Brazilian Centre for Research in Energy and Materials (CNPEM), a private non-profit organization under the supervision of the Brazilian Ministry for Science, Technology, and Innovations (MCTI), for sample preparation and characterization – LNNano/CNPEM (Proposals: 20233841, 20240216 and 20240039) and LAM (Proposals: 20231294 and 20240497) at LNLS/CNPEM, besides Marcelo R. Piton and Cilene Labre for the experimental assistance. 
I.D.B, A.R.C. and V.L.R acknowledge the support from CNPq (309920/2021-3, 306170/2023-0, 311536/2022-0). 
K.W. and T.T. acknowledge support from the JSPS KAKENHI (Grant Numbers 21H05233 and 23H02052) and World Premier International Research Center Initiative (WPI), MEXT, Japan. All authors thank Professor Marco A. Fonseca from the Federal University of Ouro Preto for supplying the clinochlore crystal and Professor C. Trallero-Giner for the fruitful discussion.

\end{acknowledgments}






%

\clearpage
\onecolumngrid

\setcounter{figure}{0}
\renewcommand{\thefigure}{S\arabic{figure}}

\large{\textbf{{This Supporting Information includes:\newline}}

\noindent    $\bullet$ Section~S1. Insulating Behavior of Clinochlore Flakes\\
    $\bullet$ Section~S2. Characterization of the Reference Sample \\
    $\bullet$ Section~S3. Additional Voltage Sweeps in the Clinochlore Device \\
    $\bullet$ Section~S4. X$^-$/X$^0$ Intensity Ratio Hysteresis \\
    $\bullet$ Section~S5. Transfer Functions

\large{\textbf{\newline
Section~S1. Insulating Behavior of Clinochlore Flakes\newline
}}

\normalsize{
An important property to be considered in our 1L-MoSe$_2$/clinochlore devices is the dielectric breakdown of clinochlore crystals. This electrical feature was investigated using parallel plane capacitors forming a gold(Au)/clinochlore/gold(Au) capacitor (see inset of Figure~S1). Here we apply a potential difference through clinochlore crystals, with different thicknesses, and we measure the maximum potential before the dielectric breakdown (V$_{BD}$). Figure~S1 brings three current \textit{versus} voltage curves for different clinochlore flakes measured in forward and backward conditions. The curves demonstrate the high-insulating behavior of clinochlore crystals, independently of the direction of the electric field. Moreover, it shows that no hysteresis is observed in the bias loop (backward \textit{versus} forward data). Therefore, these results indicate that we can neglect any significant charge tunneling in our 1L-MoSe$_2$/clinochlore device, once the \textit{V} bias range applied in the PL experiments is much smaller than the \textit{V} bias window expected to show significant charge transfer. Moreover, this observation also eliminates the possibility of a hysteresis effect induced only by the bias \textit{V}.
}

\begin{figure}[htb!]
    \centering
    \includegraphics[width=0.7\linewidth]{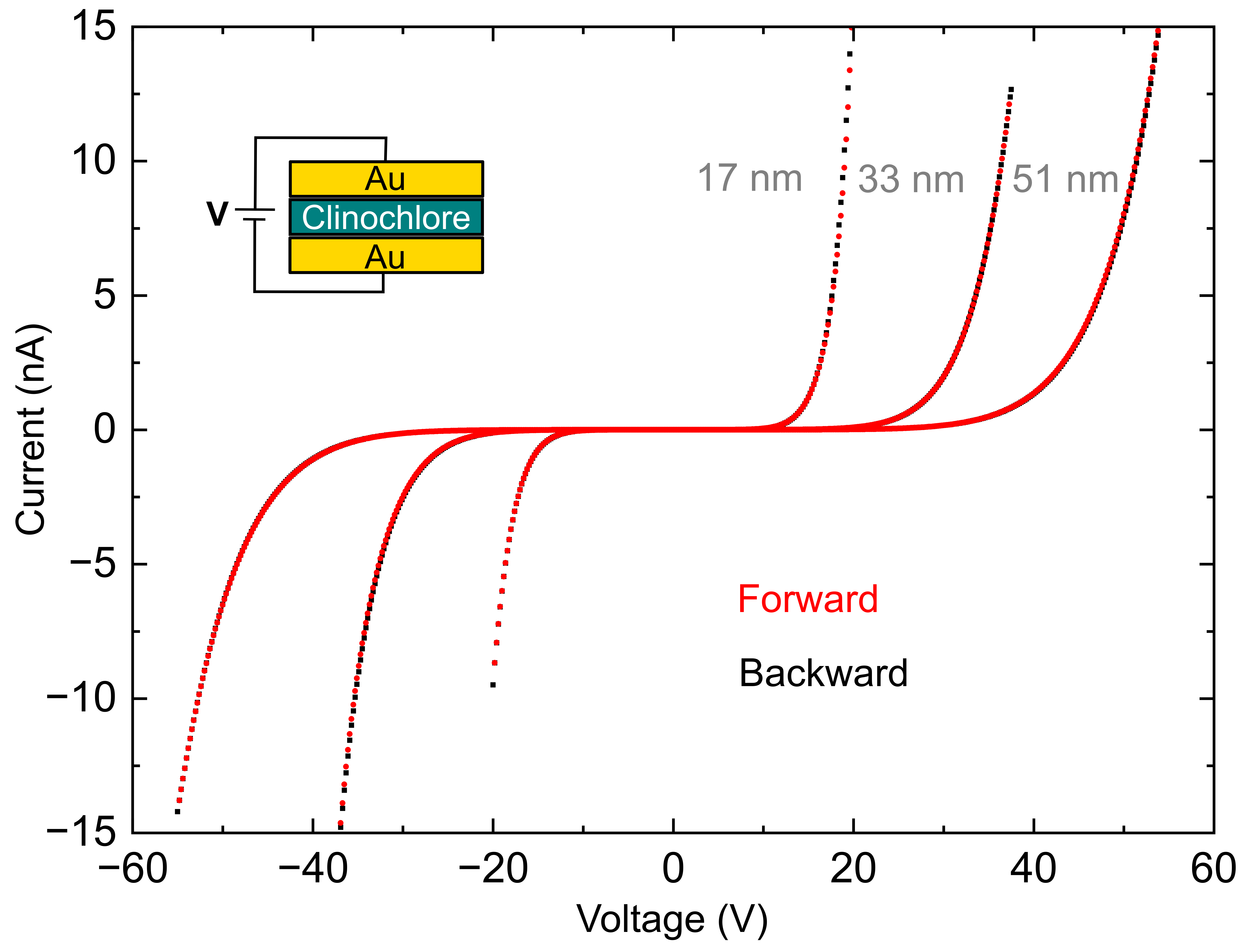}
    \caption{\textbf{Electrical breakdown of clinochlore flakes.}
    \textbf{a} Current versus Voltage bias at three representative clinochlore flakes with thicknesses of 17 nm, 33 nm, and 51 nm during forward (red) and backward (black) sweep.}
    \label{fig-BD_SI}
\end{figure}


\clearpage
\newpage
\large{\textbf{
Section~S2. Characterization of the Reference Device\newline
}}

\normalsize{
To gain deeper insights into the role of the clinochlore substrate in the optical memory effect reported in the manuscript, we also investigated a reference 1L-MoSe$_2$/hBN device. Figure~S2a shows a schematic view of our reference device in a capacitor-like structure, in which the 1L-MoSe$_2$ is sandwiched between thin hBN flakes. Figure~S2b presents the PL spectrum of the 1L-MoSe$_2$ obtained from the reference sample at 3.6~K and 0~V, exhibiting X$^0$ and X$^-$ emissions. The X$^0$ (X$^-$) peak is centered at 1.644~eV (1.617~eV) and displays a FWHM of 4.3~meV (4.0~meV). We also acquired 1L-MoSe$_2$ PL spectra from the reference sample by sweeping the gate voltage in a single cycle (0~V → V${\mathrm{max}}$ → $-$V${\mathrm{max}}$ → 0~V). Figures~S2c-f show the intensity and energy of X$^0$ and X$^-$ emissions as a function of the gate voltage. A negligible hysteresis is observed for the reference device, corroborating the fundamental contribution of the clinochlore layers for the optical memory effect.
}

\begin{figure}[htb!]
    \centering
    \includegraphics[scale=1.2]{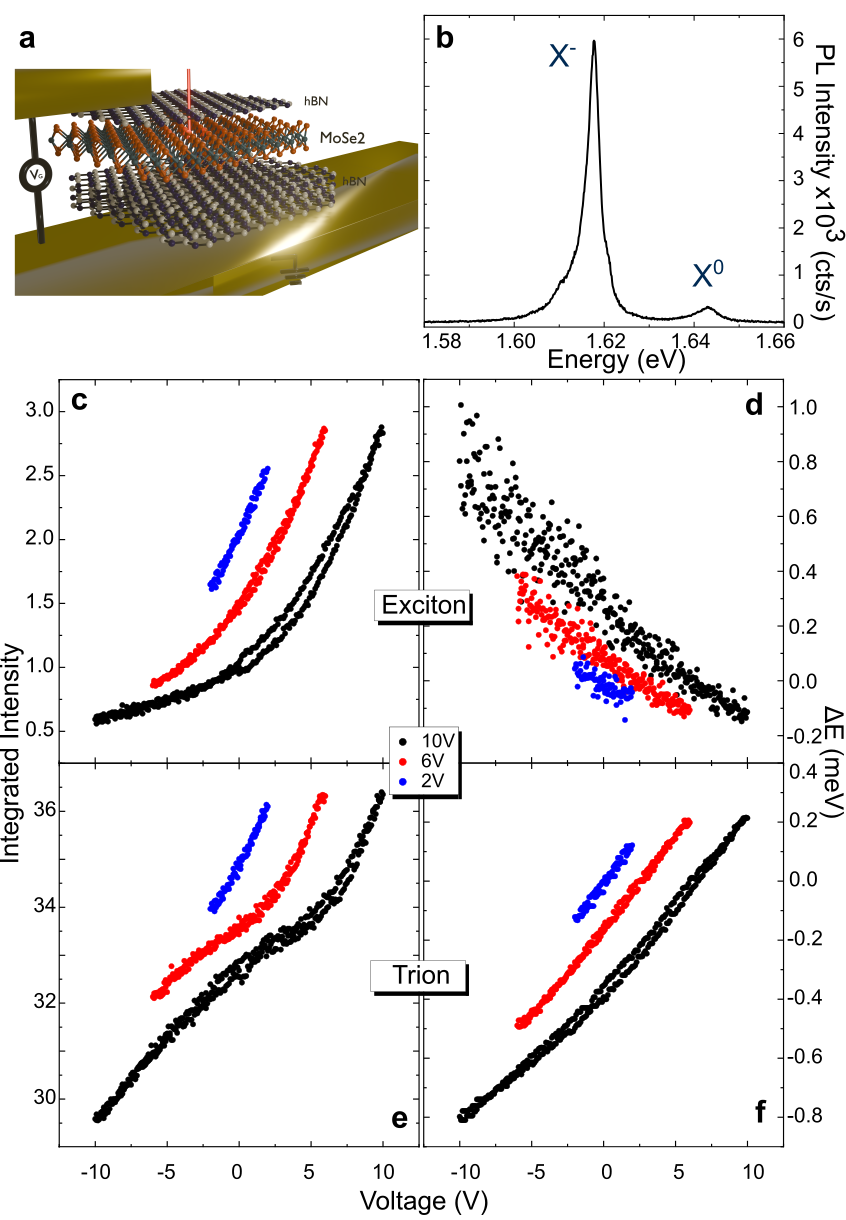}
    \caption{\textbf{Reference MoSe$_2$ device characterization.}
    \textbf{a} Schematic view of our hBN/1L-MoSe$_2$/hBN/Au-electrode device.
    \textbf{b} PL spectra of 1L-MoSe$_2$ acquired in the reference sample exhibiting X$^0$ and X$^-$ emissions.
    \textbf{c-f} Integrated intensity (c,e) and energy shift (d,f) of X$^0$ (c,d) and X$^-$ (e,f) emissions as a function of the gate voltage showing negligible hysteresis. The sweeps were performed for maximum voltages of 2~V (in blue), 6~V (in red), and 10~V (in black).}
    \label{fig-hBN_SI}
\end{figure}


\clearpage
\newpage
\large{\textbf{
Section~S3. Additional Voltage Sweeps in the Clinochlore Device\newline
}}

\normalsize{
To check the reproducibility of the reported optical memory response, we obtained 1L-MoSe$_2$ PL spectra from the clinochlore sample by sweeping the gate voltage in two cycles with opposite directions. Figures~S3a-d show the intensity and energy of X$^0$ and X$^-$ emissions as a function of the gate voltage. The similar hysteresis observed in the closed loop part of the sweeps confirms the reproducible aspect of the optical memory effect.
}

\begin{figure}[htb!]
    \centering
    \includegraphics[width=0.8\linewidth]{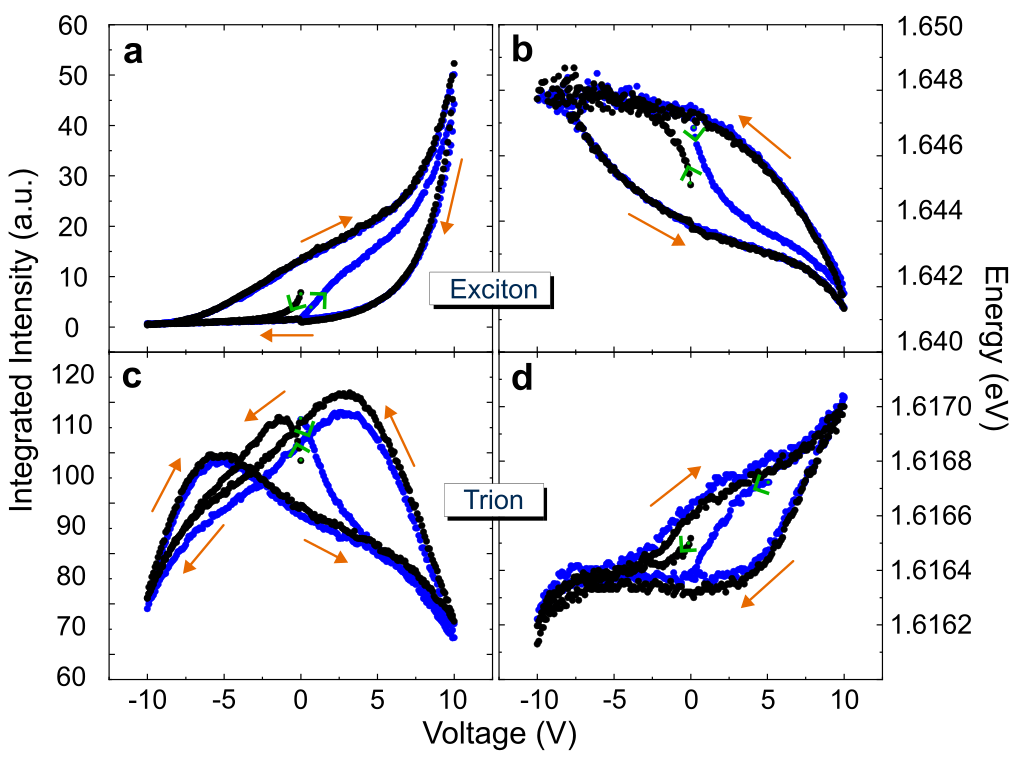}
    \caption{\textbf{Hysteresis effect in the PL emission of the clinochlore device for external voltage sweeps with opposite directions.}
    \textbf{a-d} Integrated intensity (a,c) and energy shift (b,d) of X$^0$ (a,b) and X$^-$ (c,d) emissions as a function of the gate voltage. For each graph, two voltage sweeps ranging from -10~V to 10~V are displayed. The sweeps started at 0~V and were conducted in opposite directions. The green arrows indicate where the measurement initiates and the orange arrows denote the direction of the voltage sweep.
    }
    \label{fig-Add_SI}
\end{figure}


\clearpage
\newpage
\large{\textbf{
Section~S4. X$^-$/X$^0$ Intensity Ratio Hysteresis\newline
}}

\normalsize{
As mentioned in the manuscript, hysteresis in the X$^-$/X$^0$ intensity ratio were recently reported for devices based on 1L-MoSe$_2$~\cite{choi2024tuning} and 1L-MoS$_2$~\cite{pucher2024strong} atop a perovskite substrate due to a remanent polarization. Similarly, Figure~S4 presents a robust hysteresis for the intensity ratio between X$^-$ and X$^0$ 1L-MoSe$_2$ emissions obtained from the 1L-MoSe$_2$/clinochlore device. However, while the hysteresis attributed to a remanent polarization generally exhibit a slower variation of the observables by changing the voltage sweep direction~\cite{choi2024tuning,pucher2024strong}, here we observe an abrupt modification in the X$^-$/X$^0$ intensity ratio when the sweep direction is altered. This indicates that the dynamical processes that govern the optical memory effect of the clinochlore device are more complex than a simple remanent polarization.
}

\begin{figure}[htb!]
    \centering
    \includegraphics[width=0.7\linewidth]{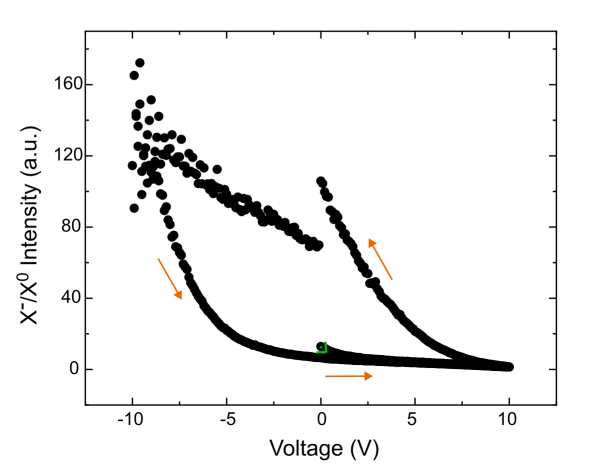}
    \caption{\textbf{X$^-$/X$^0$ intensity ratio hysteresis of the clinochlore device.}
    \textbf{a,b} Intensity ratio between X$^-$ and X$^0$ emissions as a function of the gate voltage. The green arrow indicates where the measurement initiates and the orange arrows denote the direction of the voltage sweep.}
    \label{fig-Ratio_SI}
\end{figure}


\clearpage
\newpage
\large{\textbf{
Section~S5. Transfer Functions\newline
}}

\begin{figure}[htb!]
    \centering
    \includegraphics[width=0.5\linewidth]{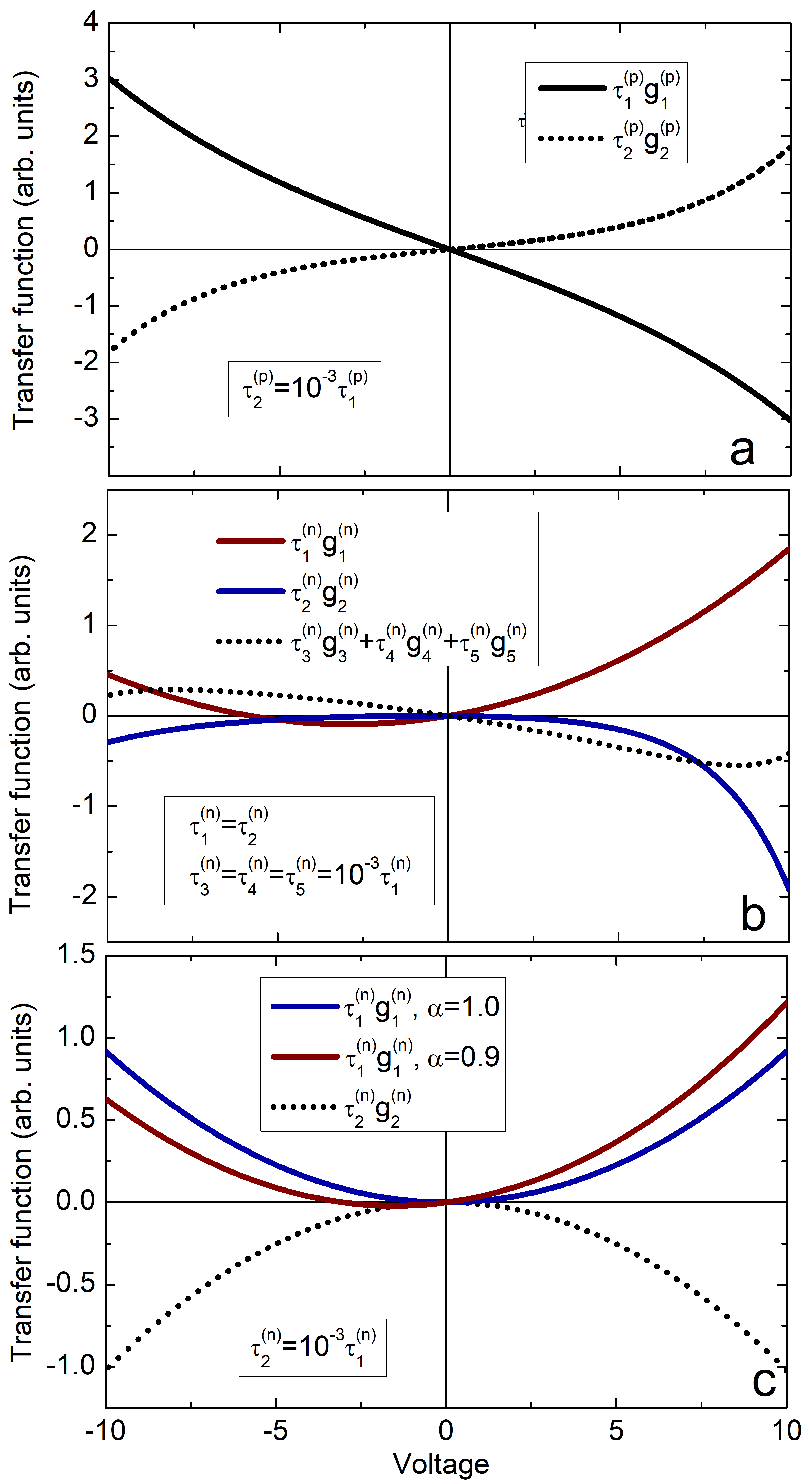}
    \caption{\textbf{Non-equilibrium transfer functions}
\textbf{a} Nonequilibrium polarization transfer functions for a leaking mechanism (solid curve) and a nonlinear polarization fluctuation (doted curve) used for the local electric field calculation displayed in panels 4 b and c of the manuscript. 
\textbf{b} Nonequilibrium charge transfer functions from the substrate to the TMD monolayer as function of applied voltage used in panel 4 d of the manuscript. 
\textbf{c} Nonequilibrium charge transfer functions from the substrate to the TMD monolayer as function of applied voltage used in panel 4 e of the manuscript. }
    \label{fig-g_SI}
\end{figure}

\normalsize{
Each independent mechanism $j$ contributing to the non-equilibrium polarization fluctuation is characterized by its relaxation time $\tau_j^{(p)}$ and the polarization transfer function $g_{j}^{(p)}(V)$, given by
\begin{equation}
g^{(p)}_j = \pm \frac{e \lambda_j}{2\eta} \left[\exp\left( \eta \frac{eV}{k_B T{eff}} \right) - \exp\left( -\eta \frac{eV}{k_B T_{eff}} \right)\right],
\label{eq2}
\end{equation}
where $e$ is the electron charge, $\lambda_j = \frac{4 \pi m^* (k_B T_{eff})^2 \exp\left( -\frac{E^{b}_j}{k_B T_{eff}} \right)}{(2 \pi \hbar)^3}$, $T_{eff}$ is the effective temperature, $E^{b}_j$ is the activation barrier for each non-equilibrium process, and $\eta < 1$ represents the local voltage efficiency drop. The positive sign in Eq.~\ref{eq2} corresponds to the contribution from charge bouncing within localization sites in the substrate, while the negative sign corresponds to the contribution from leakage \cite{WEI200564, JIN2023}. The functions used to generate the results shown in Figures~4b and 4c of the manuscript are presented in Figure~S3a.

The transfer or generation rate for charge fluctuations with relaxation time $\tau^{(n)}_j$ can be expressed as
\begin{equation}
\centering
\label{ga}
g^{(n)}_j = \pm \frac{\lambda_j A}{\eta} \left[\exp\left(\mp \eta_L \frac{eV}{k_B T}\right) + \exp\left(\pm \eta_R \frac{eV}{k_B T}\right) - 2\right],
\end{equation}
where $A$ is the device area, and
$\eta_R = \frac{\eta}{1 + \alpha}$ and $\eta_L = \frac{\eta \alpha}{1 + \alpha}$, with $\alpha \equiv \eta_L/\eta_R \in \left[0, \infty \right)$ quantifying the local symmetry break. The case of perfect symmetry, $\alpha = 1$, corresponds to $\eta_R = \eta_L = \eta/2$~\cite{LopezRichard2022}.

The transfer functions used to obtain the results in Figure~4c are presented in Figure~S3b, while the corresponding transfer functions used for Figure~4d are shown in Figure~S3c.
}

\newpage
\clearpage
\large{SUPPORTING REFERENCES\newline}

\end{document}